# Temporally resolved Type III solar radio bursts in the frequency range 3-13 MHz

Antonio Vecchio,[1,2] Milan Maksimovic,[2] Nicolina Chrysaphi,[3,4] Eduard P. Kontar,[4] and Vratislav Krupar[5,6]

[1]Radboud Radio Lab, Department of Astrophysics, Radboud University, Nijmegen, The Netherlands

[2]LESIA, Observatoire de Paris, Université PSL, CNRS, Sorbonne Université, Universit´e de Paris,5 place Jules Janssen, 92195 Meudon, France

[3]Sorbonne Université, Ecole Polytechnique, Institut Polytechnique de Paris, CNRS, Laboratoire de Physique des Plasmas (LPP), 4 Place Jussieu, 75005 Paris, France

[4]School of Physics & Astronomy, University of Glasgow,Glasgow, G12 8QQ, UK

[5]Goddard Planetary Heliophysics Institute, University of Maryland, Baltimore County, Baltimore, MD 21250, USA

[6]Heliospheric Physics Laboratory, Heliospheric Physics Division, NASA Goddard Space Flight Center, Greenbelt, MD 20771, USA

## ABSTRACT

Radio observations from space allow to characterize solar radio bursts below the ionospheric cut-off, which are otherwise inaccessible, but suffer from low, insufficient temporal resolution. In this letter we present novel, high-temporal resolution observations of Type III solar radio bursts in the range 3-13 MHz. A dedicated configuration of the RPW/HFR receiver on the Solar Orbiter mission, allowing for a temporal resolution as high as $\sim$ 0.07 s (up to two orders of magnitude better than all spacecraft measurements currently available), provides for the very first time resolved measurements of the typical decay-time values in this range of frequency. The comparison of data with different time resolution and acquired at different radial distances indicates that discrepancies with decay time values provided in previous studies are only due to the insufficient time resolution not allowing to accurately characterize decay times in this frequency range. The statistical analysis on a large sample of $\sim$ 500 type III radio bursts shows a power low decay time trend with a spectral index $-0.75 \pm 0.03$ when the median value for each frequency are considered. When these results are combined with previous observations, referring to frequencies outside the considered range, a spectral index of $-1.00 \pm 0.01$ is found in the range $\sim$ 0.1-300 MHz compatible with the presence of radio wave scattering between 1 and 100 $R_\odot$.

## 1. INTRODUCTION

Solar type III radio bursts are among the most common impulsive radio emission in the universe. They are characterized by a rapid drift in time towards lower frequencies (f ). The commonly accepted mechanism for the generation of type III radio bursts includes nonlinear wave coupling involving Langmuir waves generated by a "bump-on-tail instability" triggered by energetic electron beams. These are produced at the Sun during a flare and propagate through the plasma of the corona and the interplanetary medium with subluminal speed, typically 0.3-0.1c (where c is the light speed). As the electrons move away from the Sun they encounter a decreasing electron density $n_e$, and the Langmuir waves, emitted at the local plasma frequency $f_{pe}$ proportional to $n_e$, generate radio waves at progressively decreasing frequencies as a function of the heliocentric distance. Type III radio bursts are observed over a wide range of frequencies ranging from about $\sim$ 500 MHz down to tens of kHz close to 1 au, corresponding to a wide range of heliocentric distances. The nonlinear coupling between Langmuir waves produces electromagnetic radiation near $f_{pe}$, known as fundamental emission, or $2f_{pe}$, referred to as harmonic emission (see, e.g. Suzuki & Dulk 1985; Pick & Vilmer 2008, for reviews).

Density fluctuations along the path of the solar radio waves can strongly affect the propagation and the properties of the detected type III bursts. Scattering of radio waves on random density irregularities has long been recognized

Corresponding author: Antonio Vecchio a.vecchio@astro.ru.nl

as an important process for the interpretation of radio source sizes (e.g. Steinberg et al. 1971), positions (e.g. Fokker 1965; Stewart 1972), directivity (e.g. Thejappa et al. 2007; Bonnin et al. 2008; Reiner et al. 2009), and intensity-time profiles (e.g. Krupar et al. 2018). In particular, due to the scattering, the intensity-time profiles for a fixed frequency (hereafter "light curves") show typical features. i) The light curve of a radio burst is characterized by a very fast rising phase followed by a long-lasting exponential decrease (e.g. Evans et al. 1973). Although at frequencies greater than 100 MHz the exponential decay can be also attributed to the emission process (e.g., Li et al. 2008) this is not the case for frequencies of the order of 10 MHz (Ratcliffe et al. 2014). At lower frequencies, indeed, the behavior of the decay phase is explained as the effect of the scattering of the radio waves by electron density inhomogeneities (see e.g., Zank et al. 2024)as they propagate from the source to the detector (Krupar et al. 2018; Kontar et al. 2019). The contributions to the rise phase are less clear, although radio scattering seems to play an important role (Chrysaphi et al. 2024). ii)Since propagation effects are stronger when the frequency of the radio waves is close to the local $f_{pe}$ (e.g. Steinberg et al. 1971), lower frequencies scatter more than higher frequencies, thanks to the density scale heights being larger at larger heliocentric distances. This means that low-frequency radio photons, which undergo stronger scattering, disperse more compared to higher-frequency waves, giving rise to broader light curves with longer decay phases. This explains the relationship of inverse proportionality between decay time and frequency observed in all the data (e.g. Alexander et al. 1969; Alvarez & Haddock 1973; Barrow & Achong 1975; Krupar et al. 2018; Reid & Kontar 2018; Kontar et al. 2019; Chrysaphi et al. 2024).

Since decay times are directly related to the radio-wave scattering, their analysis and characterization can provide useful information about the strength and anisotropy of the scattering process (Kontar et al. 2019; Kuznetsov et al. 2020; Chen et al. 2020) and the level of density fluctuations (e.g. Krupar et al. 2018, 2020). By comparing radio burst properties over a large range of frequencies with advanced radio-wave propagation simulations, Kontar et al. (2019) showed that anisotropic scattering is required to reproduce the radio burst observations (decay time and source size), with the strongest scattering occurring in the perpendicular to the magnetic field direction, a result confirmed by multiple other studies (Kuznetsov et al. 2020; Chen et al. 2020; Musset et al. 2021; Kontar et al. 2023).

The plot of the decay time as a function of frequency, over the range $\sim$ 0.1-300 MHz (Kontar et al. 2019), obtained by combining various observations, shows two relevant features:

- the best-fit of the decay time $\tau$ as a function of frequency f in MHz, assuming a $f^{-\beta}$ dependence is:

$$\tau = (72.2 \pm 0.3) f^{(-0.97 \pm 0.003)} \; ; \tag{1}$$

- a data gap is present between 3 and 13 MHz due to the lack of temporally-resolved measurements. This gap represents a clear separation between measurements made from space and those made by radio observatories on ground.
  Studies from the '70s (Hartz 1964a; Boischot 1967) measured type III decay times in the 2-10 MHz (from the Alouette I spacecraft, time resolution 18 s (Hartz 1964b)) and 8-36 MHz (from the sweep frequency interferometer at the High Altitude Observatory Boulder, time resolution 4 s (Boischot 1960)) finding a dependency $\tau \propto f^{-0.98}$ and $f^{-0.8}$, respectively. More recent studies used Parker Solar Probe (PSP) data found different spectral indices: Krupar et al. (2020) found $\beta$ = 0.5 s in the frequency range 0.5-10 MHz while Jebaraj et al. (2023) detected $\beta$ = 0.73 s in the range $\sim$ 1-20 MHz when a 3.5 s time resolution is set. We note that Jebaraj et al. (2023) used a exponentially modified Gaussian fit to derive smoother type III profiles and calculate the decay time. Since the expected decay time in the frequency range considered here is of the order of 1-10 s, measurements with a time resolution of less than 1 s are therefore needed to accurately resolve the decay phase from the type III profiles. It is thus interesting to investigate whether the spread of the power law exponents in previous studies could be due to the insufficient temporal resolution of these measurements.

The range 3 and 13 MHz, roughly corresponding to the radial belt 2-5 $R_\odot$ (according to the density model by Sittler & Guhathakurta (1999)) represents a sort of boundary layer for several processes occurring in the interplanetary space. Firstly, in this range of radial distances the solar wind becomes supersonic behind the sonic critical point (Cranmer et al. 2023). At the same time, beyond about 5-7$R_\odot$ the coronal plasma density varies as $r^{-2}$, while its decrease is faster below it (see e.g Sittler & Guhathakurta 1999). Finally, the median radio flux density of type III radio bursts shows a maximum around 2 MHz, in the radial range 2-10 $R_\odot$, whose explanation is still to be found (Krupar et al. 2014; Sasikumar Raja et al. 2022). It is clear that filling the gap between 3 and 13 MHz with temporally-resolved measurements is important to better understand the dynamics of this region of interplanetary space. In particular, accurate decay time measurements in this frequency range are needed to confirm the expected trend and characterize through observational data the scattering in the radial distance belt between 2 and 5 $R_\odot$ that currently remains unexplored.

From the instrumental point of view, the requirements for this kind of measurement are conceptually easy to achieve:

- these measurements have to be made from space as the Earth's ionosphere partially reflects and absorbs the signals in this low-frequency range

- a relatively high time resolution is needed to properly sample signals with expected decay times of the order of 1-10 s.

The second point is not always easy to achieve since instruments on spacecraft are exploited for the widest possible variety of studies, and more general configurations that enable broad spectrum measurements are generally preferred. Five spacecraft currently orbiting in the interplanetary space would in principle be able to perform radio measurements in the frequency range of interest (the name of the receiver is in parentheses): Solar Orbiter (RPW), Parker Solar Probe (FIELDS), STEREO-A (SWAVES), WIND (WAVES), CHANG'E4-Queqiao (NCLE). Although the receivers of these missions are highly configurable, nominal configurations, that usually involve measurements over much wider frequency ranges to maximize the science impact and over longer integration times to minimize noise, are characterized by time resolutions of the order of seconds. It is clear that a dedicated configuration, allowing an accurate measurement of the decay times in the 3-13 MHz range is necessary.

In this letter we present for the first time, high time resolution measurements of type III burst decay times in the 3-13 MHz range by the High Frequency Receiver (HFR, Vecchio et al. 2021; Maksimovic et al. 2021) of the Radio and Plasma Waves (RPW, Maksimovic et al. 2020) instrument on board Solar Orbiter (SO, Müller et al. 2020; Zouganelis et al. 2020). These measurements, covering about 13 months of observation, were obtained by a dedicated configuration of the HFR receiver that allowed acquisition of a light curve with a sample time of 0.07 s with an average value of ~ 0.18 s. The achieved sample time is ~ 50 times better than what PSP obtained during the sixth close encounter (enhanced temporal resolution of 3.5 s) and ~ 200 times better than STEREO and WIND spacecraft.

## 2. OBSERVATION STRATEGY AND DATA

HFR is a sweeping receiver providing electric power spectral densities from 375 kHz up to 16 MHz with a maximum number of 321 frequency bins. The spectral properties of the measured signals are computed onboard where the voltage power spectral density is sampled and transmitted on ground. HFR works only in the dipole mode, namely the potential difference between couples of antennas is measured. A large variety of configurations, characterized by frequency range, number of frequency bins, temporal and spectral resolutions, etc., are programmable in a series of operating modes optimized for specific analyses (Maksimovic et al. 2020). In its nominal configurations HFR makes measurements on 50 frequency bins. In the nominal burst mode, the one between the two nominal modes with the highest time resolution, this corresponds to a time sampling of each individual HFR spectrum of about 6 seconds. It is clear that such a sampling time is not enough to carry out the decay time measurements described above, and an ad hoc configuration is needed. For this purpose, starting from 2022 December 20, HFR was initially configured to acquire five frequency bins [3.225, 5.075, 6.875, 10.025, 12.225] MHz for ten times followed by a sweep on 50 frequency bins on the full HFR band between 0.425 and 16.325 MHz including the five frequencies above. Unfortunately, after a few rounds of measurements it was realized that the bins at 5.075 and 10.025 MHz are extremely noisy, due to interference from the platform (Maksimovic et al. 2021), and do not allow for a proper measurement of the light curves of Type III bursts. It was therefore decided to replace these frequencies with 5.225 and 10.125 MHz. The final set of five frequency bins, effective from 2023 February 28, is [3.225, 5.225, 6.875, 10.125, 12.225] MHz. Although not optimal in terms of temporal resolution, the choice to have a long sweep on 50 frequencies is related to the possibility of obtaining spectrograms with a large enough number of points to produce daily dynamic spectra with sufficient resolution. The chosen HFR configuration, including performing an average on the lowest possible number of spectra (16) and measuring at only one sensor V1-V2, allows to reach, for each of the five chosen frequencies, a time resolution of ~ 0.07 s for ten times followed by an acquisition between ~ 0.5 and ~ 0.8 s, according to the frequency, due to the long sweep. Figure A.1 in Appendix A, shows an example of the HFR frequency sweep as a function of time for the receiver configuration described above.

In this study the HFR power spectral densities $V_{z}^{2}(t)$ from calibrated level 2 (L2) data were considered from 2022 December 20 to 2024 January 21 (the time interval when the HFR configuration described above was kept active).

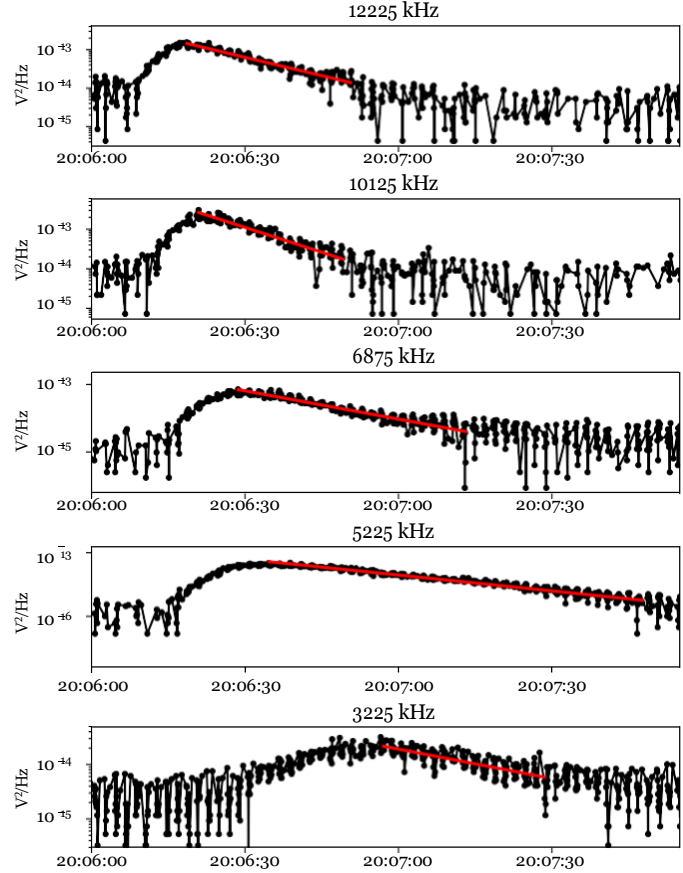

Figure 1. Light curves of the radio flux density measured by SO/RPW/HFR at five frequency channels during a type III burst on 2023 November 26. Red lines show the results of the decay time fitting.

Although L2 data are only calibrated at the receiver level ($V_f^2(t)$ is expressed in physical units $V^2 Hz^{-1}$) this does not affect the calculation of the decay time, which, at a fixed frequency, only depends on the relative amplitudes at each time that do not change when the effects of antennas are included. The rising and decay phases of the type III profile are clearly visible in Figure 1 where the light curves at four frequencies are shown for a type III burst observed by HFR. Further examples are shown in Appendix A (figures A.2, A.3, A.4).

In the analyzed dataset, obtained on a daily basis, type III bursts are identified by looking at the light curves for the five considered frequencies. Firstly the occurrence of a peak in the spectrum, possibly indicating a type III burst, is visually checked at the lowest frequency 3.225 MHz. The same peak is automatically sought at the other four frequencies in a time window 60 s before and 150 s after the peak at 3.225 MHz. The decay time $\tau$ is then obtained, for each frequency, by fitting the data points through an exponential function (Alexander et al. 1969; Aubier & Boischot 1972; Alvarez & Haddock 1973; Barrow & Achong 1975; Krupar et al. 2018; Vecchio et al. 2021) of the form

$$S(t) = S_{peak} \exp{\frac{t_{peak} - t}{\tau}}, \qquad (2)$$

after removing the background. The left and right limits of the fitting interval are chosen as the point corresponding to the 0.95 of the maximum and the last value above the background, calculated as the median value of intensity over daily data. Although the necessity to evaluate $\tau$ with functions other than a single exponential has recently been demonstrated (Chrysaphi et al. 2024), in this paper we preferred to use the classical approach to make a consistent comparison with the results found in the literature. An example of the result of the fitting procedure is shown in Figure 1.

## 3. RESULTS AND DISCUSSION

By routinely analyzing the considered HFR daily data we were able to characterize the decay time for several events at the five recorded frequencies. The number of considered events for each frequency are displayed in table A1. Decay time values were only selected according to the fit goodness: only τ values coming out of converging fits with associated R² values greater than 0.85 were considered. Figure 2(a) shows the histograms of τ for the five considered frequencies. The statistical distributions confirm that, on average, τ increases with decreasing frequency. This is a well-known property of the Type III decay time due to the greater effectiveness of the radio wave scattering at lower frequencies. The distributions of τ appear skewed with quite long tails more pronounced for lower frequencies. The origin of the tail in the distribution lies in the blind selection of events only made according to the goodness of the fit. At this stage, no visual selection was made to remove, for example, events characterized by multiple peaks for which the exponential fit, although converging from a purely mathematical point of view, has no clear physical meaning and can give rise to higher τ values.

The properties of the distributions, median and confidence interval, provide statistically relevant information about the typical values of τ at each frequency. Moreover, to have an independent estimation for each frequency, the probability density function, $P_{emp}(\tau)$, was empirically estimated using the Kernel Density Estimation (KDE, see Rosenblatt 1956; Parzen 1962, for details) providing smooth and continue functions (figure 2(a) red line) by using a suitable kernel (Silverman 1986; Hall et al. 1992; Scott 1992). Gaussian kernels were considered while the bandwidth has been automatically detected through "Scott rule of thumb" (Scott 1992), that works well for unimodal distribution. Figure 2(a) shows a very good agreement between the KDE curves and the decay time histogram.

The median value from each distribution together with the 5° and 95° percentiles are shown in table A1. The value of τ corresponding to the maximum of each $P_{emp}(\tau)$, used as another estimator of the characteristic τ, is also indicated in table A1. Figure 3(a) shows the median values from the observed distributions as a function of the frequency. These values are well in agreement with the power law (equation 1) derived by fitting Type III burst decay time measurements from both ground- and space-based data (Kontar et al. 2019).

To check the validity of the blind selection of events, the same analysis has been applied to a subset of visually selected light curves corresponding to well isolated type IIIs showing only a single peak. Table A3 shows the time of occurrence of the emission maximum, frequency of detection and radial position of the of selected type III bursts. The outcomes of this analysis have been compared to the results of weighted fits (using the intensity as weight) on all the type III burst profiles in the subset. The results are shown in figures 2(b) and Table A2. Histograms from the subset appear less skewed with respect to the entire data set. This confirms that the fits on profiles with multiple peaks provide higher values of τ contributing to the longer tail of the histograms of the total data set. Comparison of results from weighted and unweighted fits from the selected subset does not show significant differences on the histograms' shape and median. Median, confidence interval, and KDE maxima from the selected subset are slightly lower than the ones from the full dataset due the higher degree of symmetry of the corresponding histograms. For the median tau values in Table A2, the weighted values are systematically slightly larger than the unweighted ones. This effect is due to the use of the intensity as a weight for the fit. Since higher intensity values close the maximum weigh more, they produce steeper fits than the unweighted case and larger tau values are produced. The obtained results are robust with respect to the removal of noise, that is mainly due to the smaller number of photons collected because of the shorter acquisition time (see Figure 1). This was checked by applying two different filters a median and a Wiener filter, to remove the noise from the data.

Figure 3(a) shows the median decay time times for the five considered frequencies. Despite small differences, median values obtained for all three cases discussed above are in agreement with equation 1 (Figure 3(a)) showing R² values, obtained by comparing the model and the dependent variables when the linearized quantities are considered, of 0.76 (full dataset) and 0.9 (for both weighted and unweighted datasets).

Figure 3(b) shows the median values derived by the weighted fits on the selected data subset, superimposed on the collection of Type III burst decay time measurements presented by Kontar et al. (2019) and the data from Chrysaphi et al. (2024). The power law fit has been revised by including the new RPW/HFR data discussed in this letter, and the measurements of Chrysaphi et al. (2024). The new best fit function is:

$$\tau = (85.63 \pm 0.01) f^{(-1.00 \pm 0.01)} \tag{3}$$

No significant changes with respect to equation 1 (Kontar et al. 2019) were found, and the two function can be considered equal within the limits of uncertainty. We note that the spectral index β = −1 does not change if all the detected τ, instead of the median values, are used for the fit. The addition of the new observations to the dataset illustrated in Kontar et al. (2019) confirms that the dynamics of the radio wave scattering in the 2-5 $R_\odot$ radial distances belt is the one expected from the trend obtained by measurements at higher and lower frequencies, and that expected from numerical simulations of anisotropic scattering.

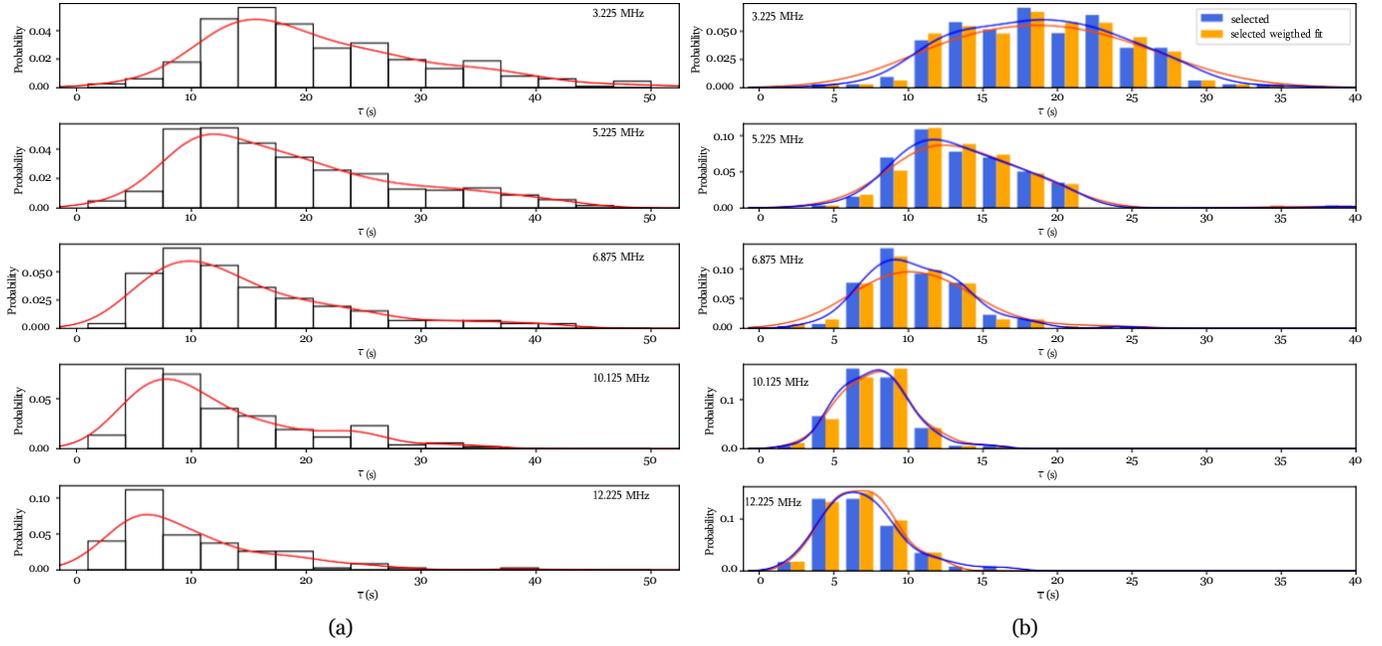

(a)                                                              (b)

Figure 2. Histograms of the decay times obtained from the full dataset (a) and from the visually selected type III burst subset (b) for unweighted (blue) and weighted (orange) fits, at the five considered frequencies. Red (a), blue and orange (b) full lines represent the empirical probability density function, $P_{emp}(\tau)$, estimated using the KDE approach and constructed from the histogram.

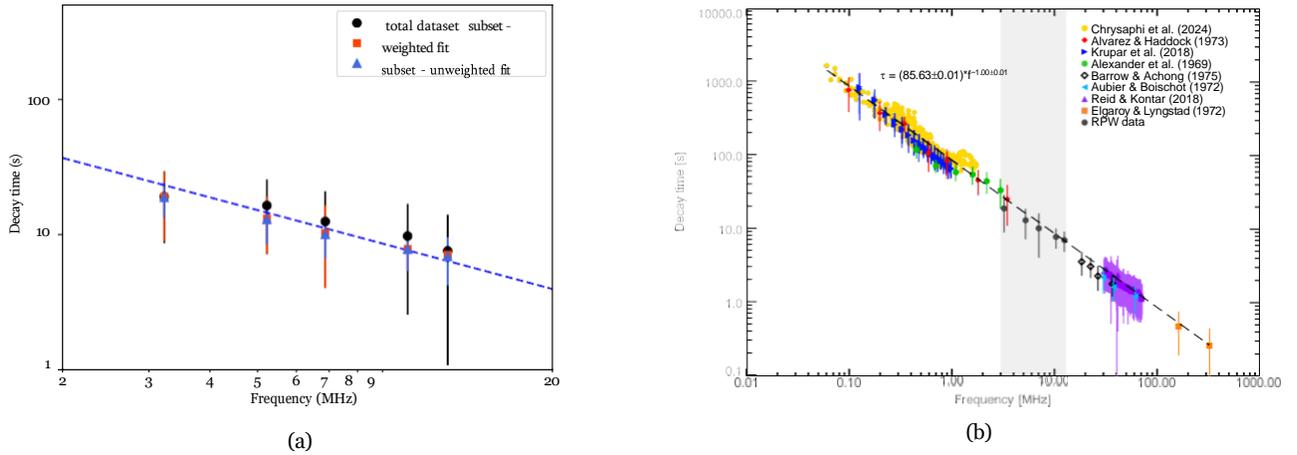

(a)                                                              (b)

Figure 3. (a) Decay times for the five considered frequencies obtained as the median value from the observed distributions. Error bars represents the standard error. The dashed line shows the power-law function from Equation (1). (b) Median decay- time values from the weighted fits on the type III bursts from the selected subset in the frequency range of 3-13 MHz (shaded region), superimposed on the data shown in Figure 10 of Kontar et al. (2019) and the newly-added data from Chrysaphi et al. (2024). Error bars represent the standard error obtained from the observed distributions. The new best fit function is also printed.

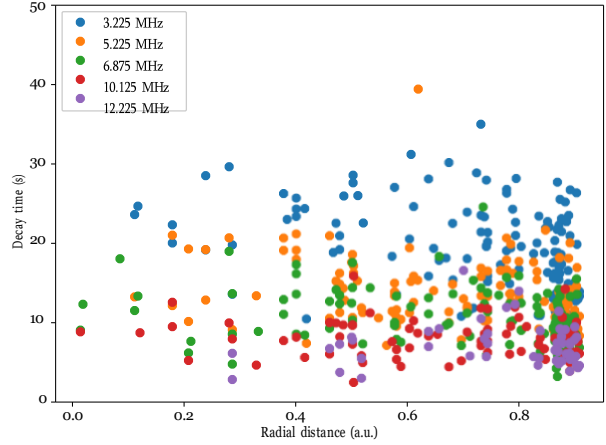

Figure 4. Decay time as a function of the radial distance for the five considered frequencies

Unlike the peak intensity of radio emission showing a maximum at ∼ 2 MHz, decay times do not show any particular feature at the same frequencies.

### 3.1. Checking the effects of radial distance and time resolution on the decay time measurement

In this section we discuss whether our results can be affected by different plasma properties, due to different radial distances, and how the τ's spectral index is depending by the time resolution of the data.

Figure 4 shows the measured decay times as a function of the radial distance of SO for the five considered frequencies. Radial distance does not correlate with decay time, consistently with the results of Chrysaphi et al. (2024). This indicates that the results shown in this letter are robust with respect to the different radial distances of the measurements in the data sample.

To study the effect of resolution on the measured decay times we reduced the time resolution of each light curve, in our visually selected dataset, to the best resolution obtained so far from radio antennas in space (3.5 s from PSP). The same analysis above has been done on the reduced resolution dataset. Histograms of the decay times from the new dataset are shown in figure A.5 in Appendix A.

Panel (a) of Figure 5 shows a comparison of the decay times obtained by the original and reduced time resolution dataset. This plot clearly highlights that at the lowest frequency of 3.225 MHz the results from the two datasets are almost comparable. The discrepancy between the two dataset increases, as expected, with increasing frequency since higher time resolution is needed to proper sample lower τ . Panel (b) shows a comparison of the median values of the decay times as a function of frequency for the full (red) and reduced time resolution (black) datasets. While the full resolution dataset well follows the trend of Equation (1), the reduced resolution dataset strongly deviates from this at higher frequencies and shows a diverging R² value. The power-law fit on the five points of both datasets provide an exponent of (−0.75 ± 0.03) for the original dataset and (−0.19 ± 0.14) for the reduced resolution dataset. The comparison of the decay time trends at different time resolution shows that in the frequency range 3-13 MHz the discrepancy with the Equation (1) increases when the time resolution of the dataset decreases and a flatten trend is obtained. This behavior is confirmed when synthetic type III profiles are considered (Appendix B). Figure 5(b) also shows the median values of the decay time obtained from 3.5 s resolution PSP data for the events listed in Jebaraj et al. (2023). These decay times have been calculated by fitting PSP light curves with the same methodology followed throughout this paper and described in Section 2. A spectral index β = −0.53 ± 0.03 is found. Also in this case the median values deviates from the trend of equation (1) showing a quite low R² equal to 0.46.

These results show that the time resolution of the dataset is the decisive factor for the proper measurement of decay times and the accurate determination of the law connecting τ to f in the range of frequency considered here. The decay time calculation by Krupar et al. (2020) and Jebaraj et al. (2023) are not accurate enough since they are performed with a sampling time of the order of the expected decay time.

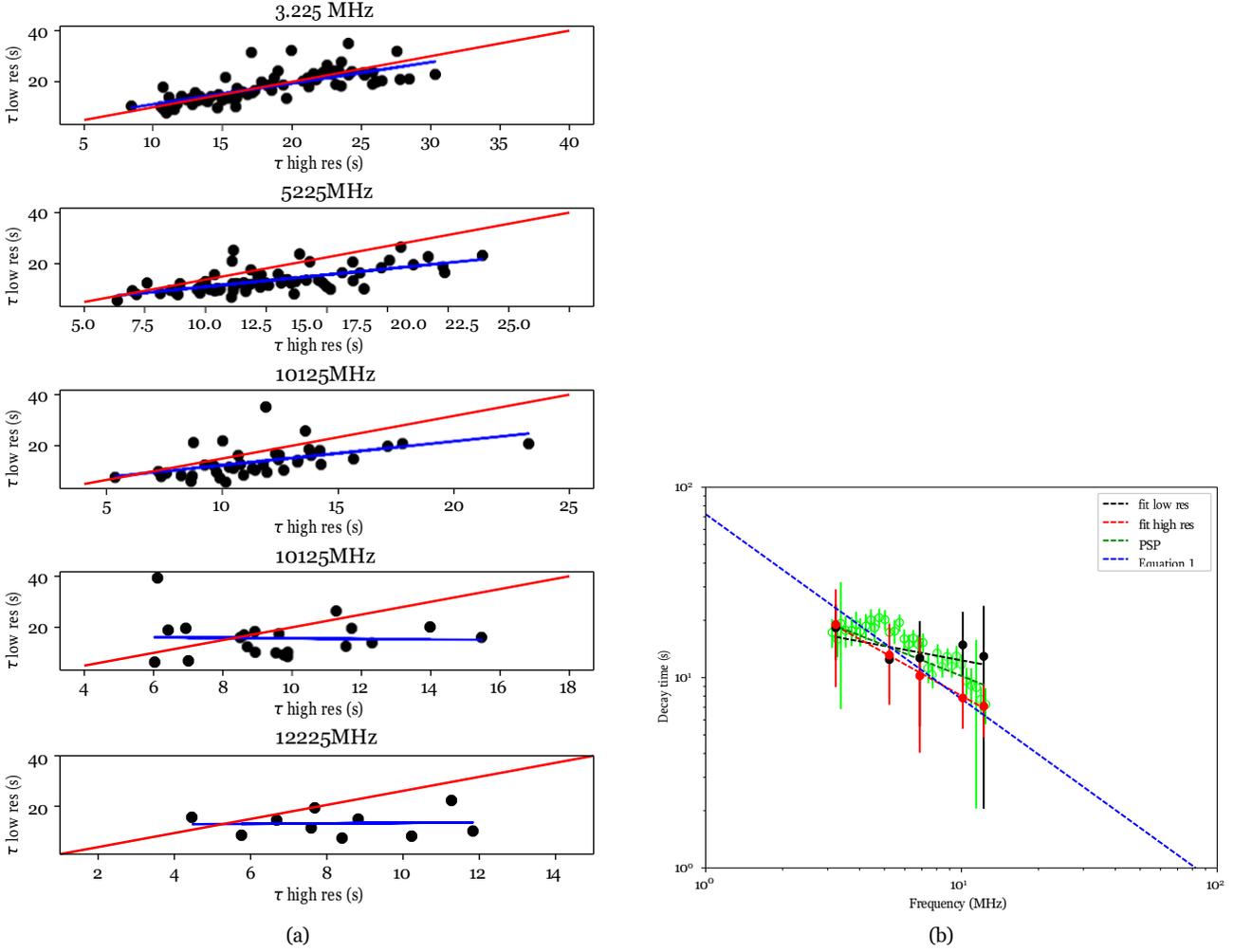

(a)                                                                    (b)

Figure 5. (a) Scatter plot of the decay times from the original dataset at higher time resolution (x-axis) and reduced resolution (y-axis) for the five considered frequencies. The blue line represents the linear fit on the data and the red line is the bisector.
(b) Decay times for the five considered frequencies obtained as the median value from the observed distributions. Red: full time resolution dataset; black: dataset with time resolution reduced to 3.5 s; green: PSP dataset with 3.5 s resolution. Error bars represent the standard error. The black, red and green dashed lines correspond to the power-law fit on the five data points. The blue dashed line shows the power-law function from Equation (1).

## 4. CONCLUSIONS

In this letter we present measurements of the Type III burst decay time from high time resolution radio data obtained using space-based instruments. The use of a special configuration of the HFR receiver of the RPW instrument onboard Solar Orbiter allowed to sample spectra over five frequencies with a temporal resolution as high as 0.07 s (with an average resolution of ∼ 0.18 s), higher than all other measurements currently available. This, together with the availability of 11 months of measurements allowed us to statistically characterize the τ behavior as a function of frequency in the range 3-13 MHz. Decay times, obtained though an exponential fit over the raw data, were calculated on the full sample of data and two sub-samples of visually selected type III profiles, corresponding to well isolated emissions whose decay time have been derived by weighted and unweighted fit. For the three datasets, the properties of the decay time distributions are in agreement with the only evident difference being the presence of skewed distributions for the full sample. This is due to the presence of events characterized by multiple peaks in the full sample for which the exponential fit can give rise to higher τ values. When the median τ is plotted as a function of frequency, a spectral index of −0.75±0.03 value is found, when a power law fit is performed, for the sub-samples of visually selected type III profiles. These results do not depend on the radial distance in a similar way as the longitudinal invariance discussed in Chrysaphi et al. (2024).

We also showed that the spectral index strongly depends on the time resolution of the radio spectra. Indeed, when the temporal resolution of the data is artificially reduced to 3.5 s, the discrepancy between median values from full and reduced resolution data increases with increasing frequency since higher time resolution is needed to proper sample lower decay times. A similar discrepancy with the full resolution RPW dataset is observed when PSP data (3.5 s resolution) are considered. The spectral index of the median decay time for the reduced resolution dataset and PSP data is $-0.19 \pm 0.14$ and $-0.53 \pm 0.03$, respectively, lower than the one from the full resolution data.

Finally we note that the decay time trend obtained from the full resolution measurements, unlike those from the reduced resolution RPW and PSP data, are compatible with the the trend of Equation (1) obtained when several decay time are fitted in the range 1-100 MHz.

All these results show that the time resolution of the dataset is the decisive factor for the proper measurement of decay times and the accurate determination of the $\tau$ spectral index in the range of frequency considered here. Discrepancies with the spectral indices obtained in previous studies (e.g. Krupar et al. 2018, 2020) do not arise by different radial distances during observations but they are due to the different time resolution of the dataset. In particular as shown in this letter, the time resolution of the measurements used in previous work is insufficient to accurately characterize the decay time for frequencies higher than 8 MHz since the sampling time is comparable to the expected decay time. On the other hand, the special and unique configuration of SO/RPW/HFR, producing measurements with the highest temporal resolution ever obtained from space in the considered range of frequency, allowed the decay time to be characterized in the range 3-13 MHz with extreme accuracy. The combination of the decay times discussed in this paper with the ones presented by Kontar et al. (2019) and Chrysaphi et al. (2024) provides a complete picture of the $\tau$ -f relationship in the range 1-100 $R_\odot$ and allow to fill the long-standing gap in fully resolved observations of type III burst decay times in the frequency range 3-13 MHz .

Solar Orbiter is a mission of international cooperation between ESA and NASA, operated by ESA. The RPW instrument has been designed and funded by CNES, CNRS, the Paris Observatory, The Swedish National Space Agency, ESA-PRODEX and all the participating institutes. N.C. acknowledges funding support from the Initiative Physique des Infinis (IPI), a research training program of the Idex SUPER at Sorbonne Universit´e. EPK was supported via the STFC/UKRI grants ST/T000422/1 and ST/Y001834/1.

## REFERENCES

Alexander, J. K., Malitson, H. H., & Stone, R. G. 1969, SoPh, 8, 388, doi: 10.1007/BF00155385

Alvarez, H., & Haddock, F. T. 1973, SoPh, 30, 175, doi: 10.1007/BF00156186

Aubier, M., & Boischot, A. 1972, A&A, 19, 343

Barrow, C. H., & Achong, A. 1975, SoPh, 45, 459, doi: 10.1007/BF00158462

Boischot, A. 1967, Annales d'Astrophysique, 30, 85

Boischot, A., Lee, R. H., & Warwick, J. W. 1960, ApJ, 131, 61, doi: 10.1086/146807

Bonnin, X., Hoang, S., & Maksimovic, M. 2008, A&A, 489, 419, doi: 10.1051/0004-6361:200809777

Chen, X., Kontar, E. P., Chrysaphi, N., et al. 2020, ApJ, 905, 43, doi: 10.3847/1538-4357/abc24e

Chrysaphi, N., Maksimovic, M., Kontar, E. P., et al. 2024, A&A, 687, L12, doi: 10.1051/0004-6361/202348175

Cranmer, S. R., Chhiber, R., Gilly, C. R., et al. 2023, SoPh, 298, 126, doi: 10.1007/s11207-023-02218-2

Evans, L. G., Fainberg, J., & Stone, R. G. 1973, SoPh, 31, 501, doi: 10.1007/BF00152825

Fokker, A. D. 1965, BAN, 18, 111

Hall, P., Marron, J. S., & Park, B. U. 1992, Probability Theory and Related Fields, 92, 1, doi: 10.1007/BF01205233

Hartz, T. R. 1964a, Annales d'Astrophysique, 27, 831

—. 1964b, Annales d'Astrophysique, 27, 823

Jebaraj, I. C., Krasnoselskikh, V., Pulupa, M., Magdalenic, J., & Bale, S. D. 2023, ApJL, 955, L20, doi: 10.3847/2041-8213/acf857

Kontar, E. P., Emslie, A. G., Clarkson, D. L., et al. 2023, ApJ, 956, 112, doi: 10.3847/1538-4357/acf6c1

Kontar, E. P., Chen, X., Chrysaphi, N., et al. 2019, ApJ, 884, 122, doi: 10.3847/1538-4357/ab40bb

Krupar, V., Maksimovic, M., Santolik, O., Cecconi, B., & Kruparova, O. 2014, SoPh, 289, 4633, doi: 10.1007/s11207-014-0601-z

Krupar, V., Santolik, O., Cecconi, B., et al. 2012, Journal of Geophysical Research (Space Physics), 117, A06101, doi: 10.1029/2011JA017333

Krupar, V., Maksimovic, M., Kontar, E. P., et al. 2018, ApJ, 857, 82, doi: 10.3847/1538-4357/aab60f

Krupar, V., Szabo, A., Maksimovic, M., et al. 2020, ApJS, 246, 57, doi: 10.3847/1538-4365/ab65bd

Kuznetsov, A. A., Chrysaphi, N., Kontar, E. P., & Motorina, G. 2020, ApJ, 898, 94, doi: 10.3847/1538-4357/aba04a

Li, B., Cairns, I. H., & Robinson, P. A. 2008, Journal of Geophysical Research (Space Physics), 113, A06104, doi: 10.1029/2007JA012957

Maksimovic, M., Bale, S. D., Chust, T., et al. 2020, A&A, 642, A12, doi: 10.1051/0004-6361/201936214

Maksimovic, M., Souˇcek, J., Chust, T., et al. 2021, A&A, 656, A41, doi: 10.1051/0004-6361/202141271

Müller, D., St. Cyr, O. C., Zouganelis, I., et al. 2020, A&A, 642, A1, doi: 10.1051/0004-6361/202038467

Musset, S., Maksimovic, M., Kontar, E., et al. 2021, A&A, Submitted

Parzen, E. 1962, The Annals of Mathematical Statistics, 33, 1065 , doi: 10.1214/aoms/1177704472

Pick, M., & Vilmer, N. 2008, A&A Rv, 16, 1, doi: 10.1007/s00159-008-0013-x

Ratcliffe, H., Kontar, E. P., & Reid, H. A. S. 2014, A&A, 572, A111, doi: 10.1051/0004-6361/201423731

Reid, H. A. S., & Kontar, E. P. 2018, A&A, 614, A69, doi: 10.1051/0004-6361/201732298

Reiner, M. J., Goetz, K., Fainberg, J., et al. 2009, SoPh, 259, 255, doi: 10.1007/s11207-009-9404-z

Rosenblatt, M. 1956, The Annals of Mathematical Statistics, 27, 832 , doi: 10.1214/aoms/1177728190

Sasikumar Raja, K., Maksimovic, M., Kontar, E. P., et al. 2022, ApJ, 924, 58, doi: 10.3847/1538-4357/ac34ed

Scott, D. W. 1992, Multivariate Density Estimation: Theory, Practice, and Visualization (New York, Chicester: John Wiley & Sons)

Silverman, B. W. 1986, Density Estimation for Statistics and Data Analysis (London: Chapman & Hall)

Sittler, Edward C., J., & Guhathakurta, M. 1999, ApJ, 523, 812, doi: 10.1086/307742

Steinberg, J. L., Aubier-Giraud, M., Leblanc, Y., & Boischot, A. 1971, A&A, 10, 362

Stewart, R. T. 1972, PASA, 2, 100, doi: 10.1017/S1323358000013059

Suzuki, S., & Dulk, G. A. 1985, in Solar Radiophysics: Studies of Emission from the Sun at Metre Wavelengths, ed. D. J. McLean & N. R. Labrum (New York: Cambridge Univ. Press), 289–332

Thejappa, G., MacDowall, R. J., & Kaiser, M. L. 2007, ApJ, 671, 894, doi: 10.1086/522664

Vecchio, A., Maksimovic, M., Krupar, V., et al. 2021, A&A, 656, A33, doi: 10.1051/0004-6361/202140988

Zank, G. P., Zhao, L. L., Adhikari, L., et al. 2024, ApJ, 966, 75, doi: 10.3847/1538-4357/ad34ab

Zouganelis, I., De Groof, A., Walsh, A. P., et al. 2020, A&A, 642, A3, doi: 10.1051/0004-6361/202038445

APPENDIX

A.   ADDITIONAL TABLES AND FIGURES

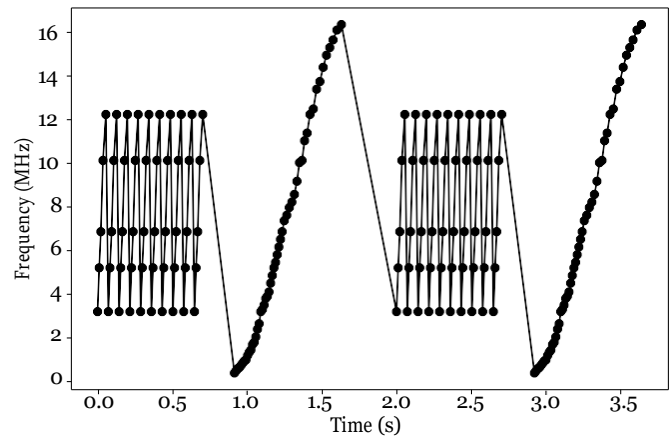

Figure A.1. Example of the HFR frequency sweep as a function of time for the receiver configuration described in the main text. Time is measured in seconds from the beginning of the sweep.

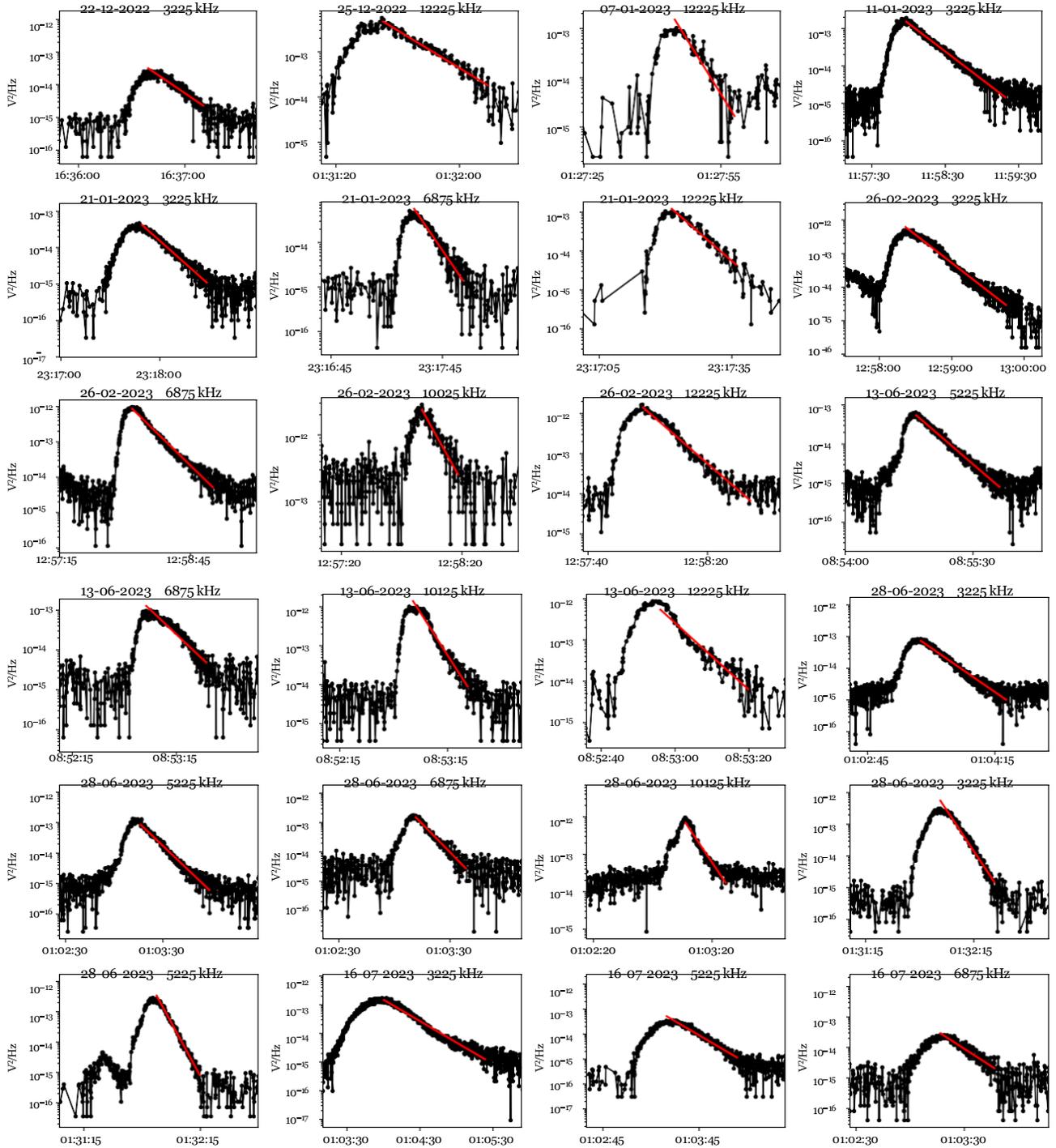

Figure A.2. Light curves of the radio flux density measured by SO/RPW/HFR during a type III burst. Red lines show the results of the decay time fitting.

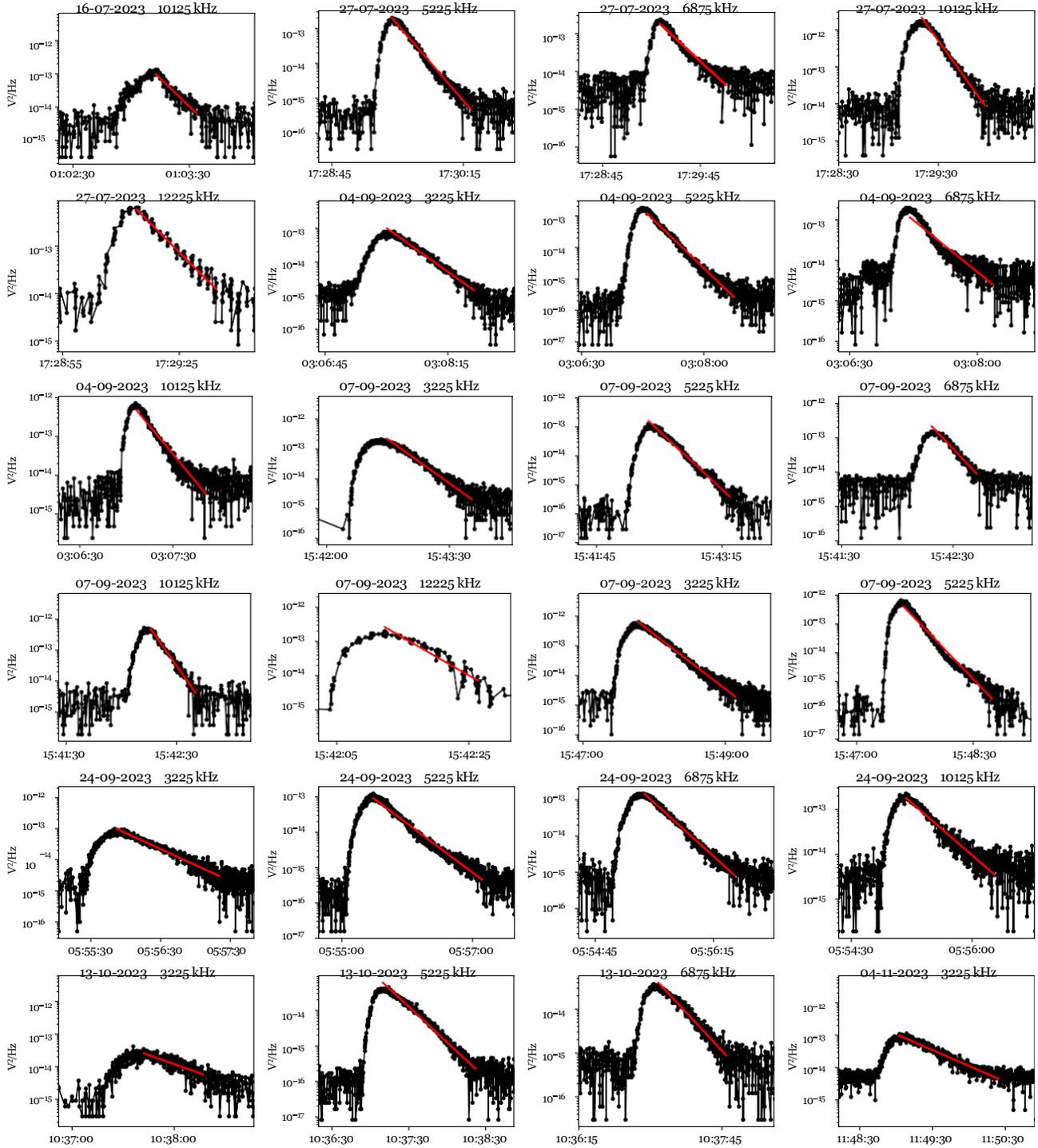

Figure A.3. Light curves of the radio flux density measured by SO/RPW/HFR during a type III burst. Red lines show the results of the decay time fitting.

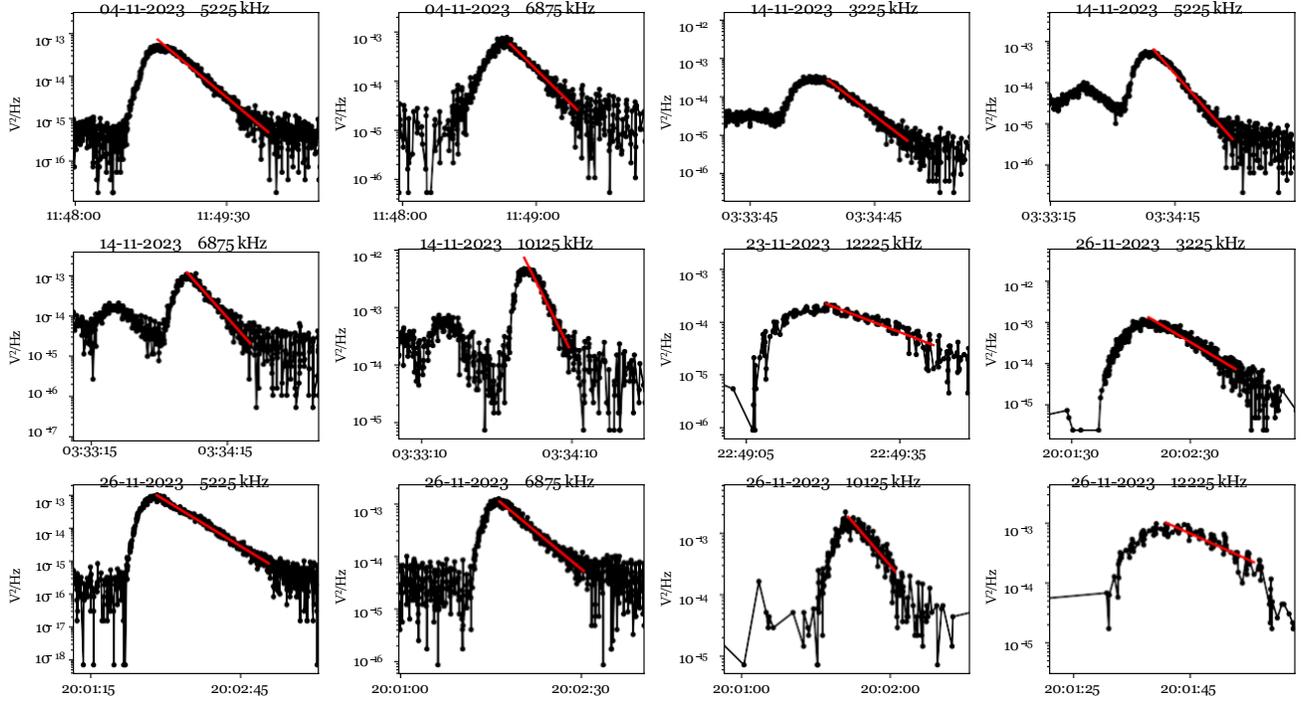

Figure A.4. Light curves of the radio flux density measured by SO/RPW/HFR during a type III burst. Red lines show the results of the decay time fitting.

| frequency (MHz) | total number of $\tau$(s) | med($\tau$)(s) | $\Delta\tau$ (s) | $\tau_{KDE}$ (s) |
|---|---|---|---|---|
| 3.225 | 349 | 19.06 | [9.98, 40.52] | 15.7 |
| 5.225 | 382 | 16.40 | [7.36, 37.07] | 11.9 |
| 6.875 | 220 | 12.49 | [5.62, 33.17] | 9.9 |
| 10.125 | 160 | 9.74 | [4.43, 25.13] | 7.75 |
| 12.225 | 107 | 7.57 | [2.60, 22.81] | 6.1 |

Table A1. Total number of measurements, from the full dataset, for each considered frequency. med($\tau$) is the median value and $\Delta\tau$ represents the 5%-95% confidence interval obtained from each distribution. $\tau_{KDE}$ corresponds to the maximum of the kernel density estimate of each distribution.

| frequency (MHz) | total number of $\tau$(s) | | med($\tau$)(s) | | $\Delta\tau$ (s) | | $\tau_{KDE}$ (s) | |
|---|---|---|---|---|---|---|---|---|
| | unweighted | weighted | unweighted | weighted | unweighted | weighted | unweighted | weighted |
| 3.225 | 135 | 137 | 18.85 | 19.01 | [10.64, 27.90] | [10.76, 28.31] | 18.92 | 18.59 |
| 5.225 | 112 | 119 | 12.95 | 13.14 | [8.01, 20.33] | [8.09, 20.52] | 11.71 | 12.35 |
| 6.875 | 113 | 117 | 10.02 | 10.23 | [5.98, 16.26] | [5.59, 18.38] | 9.18 | 10.08 |
| 10.125 | 72 | 72 | 7.78 | 7.82 | [4.53, 11.59] | [4.37, 11.98] | 8.03 | 8.04 |
| 12.225 | 50 | 49 | 6.88 | 7.07 | [3.57, 11.85] | [3.60, 10.85] | 6.30 | 6.96 |

Table A2. Results of the statistical analysis on the visually selected subset of type III bursts when both unweighted and weighted exponential fits are used. med($\tau$) is the median value and $\Delta\tau$ represents the 5%-95% confidence interval obtained from each distribution. $\tau_{KDE}$ corresponds to the maximum of the kernel density estimate of each distribution.

Table A3. Time of occurrence of the emission maximum, frequency of detection and radial position of the of the visually selected subset of type III bursts.

| date (UT) | time (UT) | Frequency (kHz) | Radial distance (au) |
|-----------|-----------|-----------------|----------------------|
| 20/12/2022 | 04:08:59 | 3225 | 0.86 |
| 20/12/2022 | 04:08:59 | 6875 | 0.86 |
| 21/12/2022 | 03:51:00 | 3225 | 0.86 |
| 21/12/2022 | 03:51:00 | 6875 | 0.86 |
| 21/12/2022 | 03:51:00 | 12225 | 0.86 |
| 21/12/2022 | 04:53:00 | 6875 | 0.86 |
| 21/12/2022 | 05:33:59 | 3225 | 0.86 |
| 22/12/2022 | 12:57:59 | 6875 | 0.86 |
| 22/12/2022 | 16:36:00 | 3225 | 0.86 |
| 22/12/2022 | 16:36:00 | 6875 | 0.86 |
| 22/12/2022 | 18:39:00 | 3225 | 0.86 |
| 22/12/2022 | 18:39:00 | 6875 | 0.86 |
| 22/12/2022 | 19:57:59 | 3225 | 0.86 |
| 22/12/2022 | 19:58:00 | 6875 | 0.86 |
| 24/12/2022 | 04:10:59 | 3225 | 0.87 |
| 24/12/2022 | 04:10:59 | 6875 | 0.87 |
| 24/12/2022 | 07:56:00 | 3225 | 0.87 |
| 24/12/2022 | 07:55:59 | 6875 | 0.87 |
| 25/12/2022 | 08:43:00 | 3225 | 0.87 |
| 25/12/2022 | 08:43:00 | 6875 | 0.87 |
| 25/12/2022 | 08:42:59 | 12225 | 0.87 |
| 27/12/2022 | 10:00:59 | 3225 | 0.87 |
| 28/12/2022 | 00:58:00 | 3225 | 0.87 |
| 28/12/2022 | 00:58:00 | 6875 | 0.87 |
| 28/12/2022 | 00:57:59 | 12225 | 0.87 |
| 28/12/2022 | 01:09:00 | 3225 | 0.87 |
| 28/12/2022 | 08:31:00 | 3225 | 0.87 |
| 01/01/2023 | 02:03:00 | 3225 | 0.87 |
| 01/01/2023 | 02:03:00 | 6875 | 0.87 |
| 01/01/2023 | 02:02:59 | 12225 | 0.87 |
| 01/01/2023 | 07:45:59 | 3225 | 0.87 |
| 02/01/2023 | 22:52:00 | 3225 | 0.87 |
| 06/01/2023 | 21:59:59 | 3225 | 0.87 |
| 06/01/2023 | 22:00:00 | 6875 | 0.87 |
| 07/01/2023 | 00:10:00 | 6875 | 0.87 |
| 07/01/2023 | 00:10:00 | 12225 | 0.87 |
| 07/01/2023 | 01:37:59 | 3225 | 0.87 |
| 07/01/2023 | 01:37:59 | 6875 | 0.87 |

**Table A3** *(continued)*

| date (UT) | time (UT) | Frequency (kHz) | Radial distance (au) |
| --- | --- | --- | --- |
| 07/01/2023 | 01:37:59 | 12225 | 0.87 |
| 07/01/2023 | 03:18:00 | 3225 | 0.87 |
| 07/01/2023 | 03:18:00 | 6875 | 0.87 |
| 08/01/2023 | 12:29:59 | 3225 | 0.86 |
| 08/01/2023 | 12:30:00 | 6875 | 0.86 |
| 08/01/2023 | 12:29:59 | 12225 | 0.86 |
| 09/01/2023 | 23:48:59 | 3225 | 0.86 |
| 09/01/2023 | 23:48:59 | 6875 | 0.86 |
| 09/01/2023 | 23:49:00 | 12225 | 0.86 |
| 11/01/2023 | 00:51:00 | 3225 | 0.86 |
| 11/01/2023 | 00:51:00 | 6875 | 0.86 |
| 11/01/2023 | 09:18:59 | 3225 | 0.86 |
| 11/01/2023 | 09:18:59 | 6875 | 0.86 |
| 11/01/2023 | 11:57:00 | 3225 | 0.86 |
| 15/01/2023 | 11:17:59 | 3225 | 0.85 |
| 16/01/2023 | 03:47:00 | 3225 | 0.85 |
| 16/01/2023 | 03:47:00 | 6875 | 0.85 |
| 16/01/2023 | 04:34:59 | 3225 | 0.85 |
| 20/01/2023 | 04:51:59 | 3225 | 0.83 |
| 20/01/2023 | 04:51:59 | 6875 | 0.83 |
| 21/01/2023 | 23:16:59 | 3225 | 0.83 |
| 21/01/2023 | 23:16:59 | 6875 | 0.83 |
| 21/01/2023 | 23:16:59 | 12225 | 0.83 |
| 30/01/2023 | 06:39:00 | 3225 | 0.79 |
| 05/02/2023 | 10:43:59 | 6875 | 0.76 |
| 09/02/2023 | 06:39:00 | 3225 | 0.74 |
| 09/02/2023 | 09:15:00 | 6875 | 0.74 |
| 09/02/2023 | 09:14:59 | 12225 | 0.74 |
| 09/02/2023 | 15:00:00 | 3225 | 0.74 |
| 09/02/2023 | 16:58:59 | 3225 | 0.74 |
| 10/02/2023 | 05:03:59 | 6875 | 0.73 |
| 12/02/2023 | 05:37:00 | 3225 | 0.72 |
| 13/02/2023 | 12:41:59 | 3225 | 0.71 |
| 13/02/2023 | 12:41:59 | 6875 | 0.71 |
| 14/02/2023 | 08:51:59 | 3225 | 0.71 |
| 14/02/2023 | 08:51:59 | 6875 | 0.71 |
| 14/02/2023 | 08:52:00 | 12225 | 0.71 |
| 19/02/2023 | 01:21:59 | 3225 | 0.68 |
| 19/02/2023 | 01:21:59 | 6875 | 0.68 |
| 21/02/2023 | 02:48:59 | 3225 | 0.66 |

**Table A3** *(continued)*

| date (UT) | time (UT) | Frequency (kHz) | Radial distance (au) |
|-----------|-----------|-----------------|----------------------|
| 21/02/2023 | 02:48:59 | 6875 | 0.66 |
| 23/02/2023 | 23:15:00 | 3225 | 0.65 |
| 23/02/2023 | 23:15:00 | 6875 | 0.65 |
| 23/02/2023 | 23:15:00 | 12225 | 0.65 |
| 26/02/2023 | 12:58:00 | 3225 | 0.63 |
| 26/02/2023 | 12:58:00 | 6875 | 0.63 |
| 26/02/2023 | 12:57:59 | 10125 | 0.63 |
| 26/02/2023 | 12:57:59 | 12225 | 0.63 |
| 26/02/2023 | 18:34:00 | 6875 | 0.63 |
| 26/02/2023 | 19:13:00 | 3225 | 0.63 |
| 26/02/2023 | 19:13:00 | 6875 | 0.63 |
| 26/02/2023 | 19:13:00 | 12225 | 0.63 |
| 01/03/2023 | 01:36:59 | 5225 | 0.61 |
| 03/03/2023 | 04:48:59 | 3225 | 0.60 |
| 03/03/2023 | 04:48:59 | 5225 | 0.60 |
| 03/03/2023 | 04:48:59 | 6875 | 0.60 |
| 18/03/2023 | 02:31:00 | 5225 | 0.50 |
| 18/03/2023 | 02:30:59 | 10125 | 0.50 |
| 18/03/2023 | 09:09:00 | 10125 | 0.50 |
| 18/03/2023 | 16:16:00 | 5225 | 0.50 |
| 20/03/2023 | 02:10:59 | 5225 | 0.48 |
| 21/03/2023 | 08:51:59 | 5225 | 0.48 |
| 21/03/2023 | 10:20:00 | 10125 | 0.48 |
| 21/03/2023 | 10:20:00 | 12225 | 0.48 |
| 23/03/2023 | 17:31:00 | 5225 | 0.46 |
| 23/03/2023 | 17:31:00 | 6875 | 0.46 |
| 23/03/2023 | 17:31:00 | 10125 | 0.46 |
| 23/03/2023 | 18:50:59 | 5225 | 0.46 |
| 23/03/2023 | 18:50:59 | 6875 | 0.46 |
| 23/03/2023 | 18:51:00 | 10125 | 0.46 |
| 23/03/2023 | 18:51:00 | 12225 | 0.46 |
| 27/03/2023 | 21:59:59 | 3225 | 0.41 |
| 27/03/2023 | 21:59:59 | 5225 | 0.41 |
| 30/03/2023 | 10:58:59 | 3225 | 0.37 |
| 30/03/2023 | 10:59:00 | 5225 | 0.37 |
| 30/03/2023 | 10:59:00 | 6875 | 0.37 |
| 30/03/2023 | 10:59:00 | 10125 | 0.37 |
| 30/03/2023 | 23:10:00 | 5225 | 0.37 |
| 30/03/2023 | 23:10:00 | 6875 | 0.37 |
| 04/04/2023 | 03:53:59 | 5225 | 0.28 |

**Table A3** *(continued)*

| date (UT) | time (UT) | Frequency (kHz) | Radial distance (au) |
|-----------|-----------|-----------------|----------------------|
| 04/04/2023 | 03:53:59 | 6875 | 0.28 |
| 04/04/2023 | 03:54:00 | 12225 | 0.28 |
| 04/04/2023 | 04:11:00 | 3225 | 0.28 |
| 04/04/2023 | 04:10:59 | 6875 | 0.28 |
| 04/04/2023 | 04:11:00 | 12225 | 0.28 |
| 04/04/2023 | 04:38:00 | 3225 | 0.28 |
| 04/04/2023 | 16:12:00 | 6875 | 0.28 |
| 04/04/2023 | 16:12:00 | 10125 | 0.28 |
| 06/04/2023 | 06:42:59 | 3225 | 0.23 |
| 06/04/2023 | 08:47:59 | 3225 | 0.23 |
| 06/04/2023 | 08:47:59 | 5225 | 0.23 |
| 06/04/2023 | 18:59:00 | 5225 | 0.23 |
| 07/04/2023 | 02:39:00 | 6875 | 0.21 |
| 03/05/2023 | 01:49:00 | 3225 | 0.47 |
| 03/05/2023 | 01:48:59 | 5225 | 0.47 |
| 03/05/2023 | 01:48:59 | 6875 | 0.47 |
| 03/05/2023 | 01:48:59 | 10125 | 0.47 |
| 31/05/2023 | 04:00:59 | 3225 | 0.78 |
| 31/05/2023 | 04:01:00 | 5225 | 0.78 |
| 31/05/2023 | 04:01:00 | 6875 | 0.78 |
| 10/06/2023 | 14:40:59 | 3225 | 0.84 |
| 10/06/2023 | 14:40:59 | 5225 | 0.84 |
| 10/06/2023 | 14:40:59 | 6875 | 0.84 |
| 10/06/2023 | 14:40:59 | 10125 | 0.84 |
| 14/06/2023 | 03:01:59 | 3225 | 0.86 |
| 14/06/2023 | 03:01:59 | 5225 | 0.86 |
| 24/06/2023 | 04:33:59 | 3225 | 0.88 |
| 25/06/2023 | 15:55:59 | 3225 | 0.88 |
| 25/06/2023 | 15:56:00 | 5225 | 0.88 |
| 27/06/2023 | 23:31:59 | 3225 | 0.88 |
| 27/06/2023 | 23:31:59 | 5225 | 0.88 |
| 28/06/2023 | 01:03:00 | 3225 | 0.88 |
| 28/06/2023 | 01:03:00 | 5225 | 0.88 |
| 28/06/2023 | 01:03:00 | 6875 | 0.88 |
| 28/06/2023 | 01:31:00 | 3225 | 0.88 |
| 28/06/2023 | 01:31:00 | 5225 | 0.88 |
| 28/06/2023 | 03:47:00 | 3225 | 0.88 |
| 28/06/2023 | 03:46:59 | 5225 | 0.88 |
| 28/06/2023 | 03:46:59 | 6875 | 0.88 |
| 28/06/2023 | 03:46:59 | 10125 | 0.88 |

**Table A3** *(continued)*

| date (UT) | time (UT) | Frequency (kHz) | Radial distance (au) |
|---|---|---|---|
| 28/06/2023 | 03:47:00 | 12225 | 0.88 |
| 28/06/2023 | 10:57:59 | 5225 | 0.88 |
| 28/06/2023 | 10:57:59 | 6875 | 0.88 |
| 28/06/2023 | 10:58:00 | 10125 | 0.88 |
| 28/06/2023 | 10:58:00 | 12225 | 0.88 |
| 29/06/2023 | 00:24:00 | 3225 | 0.89 |
| 29/06/2023 | 00:24:00 | 5225 | 0.89 |
| 29/06/2023 | 00:24:00 | 6875 | 0.89 |
| 29/06/2023 | 00:24:00 | 10125 | 0.89 |
| 29/06/2023 | 00:23:59 | 12225 | 0.89 |
| 29/06/2023 | 03:42:59 | 6875 | 0.89 |
| 29/06/2023 | 03:42:59 | 10125 | 0.89 |
| 29/06/2023 | 03:43:00 | 12225 | 0.89 |
| 01/07/2023 | 11:31:59 | 3225 | 0.89 |
| 02/07/2023 | 00:15:59 | 5225 | 0.89 |
| 02/07/2023 | 00:15:59 | 6875 | 0.89 |
| 02/07/2023 | 00:15:59 | 10125 | 0.89 |
| 02/07/2023 | 00:15:59 | 12225 | 0.89 |
| 03/07/2023 | 12:46:00 | 3225 | 0.89 |
| 03/07/2023 | 12:46:00 | 5225 | 0.89 |
| 04/07/2023 | 06:41:00 | 3225 | 0.89 |
| 04/07/2023 | 06:41:00 | 5225 | 0.89 |
| 07/07/2023 | 22:35:00 | 3225 | 0.88 |
| 07/07/2023 | 22:34:59 | 5225 | 0.88 |
| 08/07/2023 | 18:53:59 | 3225 | 0.88 |
| 08/07/2023 | 18:53:59 | 5225 | 0.88 |
| 08/07/2023 | 18:54:00 | 6875 | 0.88 |
| 08/07/2023 | 18:54:00 | 10125 | 0.88 |
| 10/07/2023 | 10:44:59 | 3225 | 0.88 |
| 10/07/2023 | 10:44:59 | 5225 | 0.88 |
| 10/07/2023 | 10:44:59 | 6875 | 0.88 |
| 10/07/2023 | 10:44:59 | 10125 | 0.88 |
| 10/07/2023 | 10:44:59 | 12225 | 0.88 |
| 12/07/2023 | 11:14:59 | 10125 | 0.88 |
| 14/07/2023 | 10:01:00 | 10125 | 0.87 |
| 14/07/2023 | 10:01:00 | 12225 | 0.87 |
| 16/07/2023 | 01:04:00 | 3225 | 0.87 |
| 16/07/2023 | 01:04:00 | 5225 | 0.87 |
| 16/07/2023 | 01:04:00 | 6875 | 0.87 |
| 16/07/2023 | 01:04:00 | 10125 | 0.87 |

**Table A3** *(continued)*

| date (UT) | time (UT) | Frequency (kHz) | Radial distance (au) |
|-----------|-----------|-----------------|----------------------|
| 16/07/2023 | 22:21:00 | 3225 | 0.87 |
| 16/07/2023 | 22:21:00 | 5225 | 0.87 |
| 16/07/2023 | 22:21:00 | 6875 | 0.87 |
| 18/07/2023 | 00:32:59 | 3225 | 0.86 |
| 18/07/2023 | 18:41:00 | 6875 | 0.86 |
| 18/07/2023 | 21:45:59 | 3225 | 0.86 |
| 18/07/2023 | 21:46:00 | 5225 | 0.86 |
| 19/07/2023 | 20:48:00 | 3225 | 0.86 |
| 19/07/2023 | 20:48:00 | 5225 | 0.86 |
| 19/07/2023 | 20:48:00 | 6875 | 0.86 |
| 19/07/2023 | 20:47:59 | 10125 | 0.86 |
| 19/07/2023 | 20:47:59 | 12225 | 0.86 |
| 20/07/2023 | 09:59:00 | 3225 | 0.86 |
| 20/07/2023 | 18:31:59 | 3225 | 0.86 |
| 20/07/2023 | 18:31:59 | 5225 | 0.86 |
| 20/07/2023 | 18:31:59 | 6875 | 0.86 |
| 24/07/2023 | 04:57:59 | 3225 | 0.84 |
| 24/07/2023 | 04:57:59 | 5225 | 0.84 |
| 24/07/2023 | 04:58:00 | 10125 | 0.84 |
| 27/07/2023 | 17:29:00 | 3225 | 0.83 |
| 27/07/2023 | 17:29:00 | 5225 | 0.83 |
| 27/07/2023 | 17:29:00 | 6875 | 0.83 |
| 27/07/2023 | 17:28:59 | 10125 | 0.83 |
| 27/07/2023 | 17:28:59 | 12225 | 0.83 |
| 03/08/2023 | 09:12:00 | 3225 | 0.80 |
| 03/08/2023 | 09:11:59 | 6875 | 0.80 |
| 03/08/2023 | 09:11:59 | 10125 | 0.80 |
| 03/08/2023 | 19:58:59 | 5225 | 0.80 |
| 03/08/2023 | 19:58:59 | 6875 | 0.80 |
| 05/08/2023 | 05:23:59 | 3225 | 0.79 |
| 05/08/2023 | 10:58:59 | 3225 | 0.79 |
| 07/08/2023 | 16:30:59 | 3225 | 0.78 |
| 07/08/2023 | 16:31:00 | 5225 | 0.78 |
| 07/08/2023 | 16:31:00 | 6875 | 0.78 |
| 07/08/2023 | 16:31:00 | 10125 | 0.78 |
| 07/08/2023 | 20:52:00 | 5225 | 0.78 |
| 07/08/2023 | 20:52:00 | 6875 | 0.78 |
| 08/08/2023 | 01:05:00 | 3225 | 0.77 |
| 08/08/2023 | 01:05:00 | 5225 | 0.77 |
| 08/08/2023 | 11:38:59 | 3225 | 0.77 |

**Table A3** *(continued)*

| date (UT) | time (UT) | Frequency (kHz) | Radial distance (au) |
|---|---|---|---|
| 08/08/2023 | 11:38:59 | 5225 | 0.77 |
| 10/08/2023 | 07:14:59 | 5225 | 0.76 |
| 10/08/2023 | 22:02:00 | 3225 | 0.76 |
| 10/08/2023 | 22:02:00 | 5225 | 0.76 |
| 14/08/2023 | 07:06:00 | 3225 | 0.74 |
| 14/08/2023 | 07:06:00 | 10125 | 0.74 |
| 14/08/2023 | 07:06:00 | 12225 | 0.74 |
| 14/08/2023 | 09:28:00 | 3225 | 0.74 |
| 14/08/2023 | 09:28:00 | 5225 | 0.74 |
| 14/08/2023 | 09:28:00 | 6875 | 0.74 |
| 14/08/2023 | 09:27:59 | 10125 | 0.74 |
| 14/08/2023 | 16:26:59 | 3225 | 0.74 |
| 14/08/2023 | 16:26:59 | 5225 | 0.74 |
| 14/08/2023 | 16:27:00 | 10125 | 0.74 |
| 14/08/2023 | 16:27:00 | 12225 | 0.74 |
| 14/08/2023 | 17:46:59 | 3225 | 0.74 |
| 14/08/2023 | 17:46:59 | 5225 | 0.74 |
| 14/08/2023 | 17:46:59 | 6875 | 0.74 |
| 14/08/2023 | 17:47:00 | 10125 | 0.74 |
| 15/08/2023 | 00:45:00 | 3225 | 0.73 |
| 15/08/2023 | 00:45:00 | 5225 | 0.73 |
| 15/08/2023 | 00:45:00 | 6875 | 0.73 |
| 15/08/2023 | 00:44:59 | 10125 | 0.73 |
| 15/08/2023 | 00:44:59 | 12225 | 0.73 |
| 16/08/2023 | 00:59:00 | 5225 | 0.73 |
| 16/08/2023 | 00:59:00 | 6875 | 0.73 |
| 16/08/2023 | 00:59:00 | 10125 | 0.73 |
| 16/08/2023 | 00:59:00 | 12225 | 0.73 |
| 16/08/2023 | 07:09:00 | 3225 | 0.73 |
| 16/08/2023 | 13:07:59 | 3225 | 0.73 |
| 16/08/2023 | 13:07:59 | 5225 | 0.73 |
| 16/08/2023 | 13:08:00 | 10125 | 0.73 |
| 16/08/2023 | 16:00:59 | 3225 | 0.73 |
| 16/08/2023 | 16:00:59 | 5225 | 0.73 |
| 16/08/2023 | 16:32:59 | 10125 | 0.73 |
| 20/08/2023 | 13:29:00 | 3225 | 0.70 |
| 20/08/2023 | 13:28:59 | 5225 | 0.70 |
| 20/08/2023 | 13:28:59 | 6875 | 0.70 |
| 20/08/2023 | 13:28:59 | 10125 | 0.70 |
| 20/08/2023 | 14:16:59 | 3225 | 0.70 |

**Table A3** *(continued)*

| date (UT) | time (UT) | Frequency (kHz) | Radial distance (au) |
|-----------|-----------|-----------------|----------------------|
| 20/08/2023 | 14:16:59 | 5225 | 0.70 |
| 21/08/2023 | 16:27:00 | 5225 | 0.70 |
| 21/08/2023 | 16:26:59 | 12225 | 0.70 |
| 24/08/2023 | 20:36:00 | 3225 | 0.68 |
| 24/08/2023 | 20:36:00 | 5225 | 0.68 |
| 25/08/2023 | 03:26:59 | 5225 | 0.67 |
| 25/08/2023 | 03:26:59 | 6875 | 0.67 |
| 25/08/2023 | 03:27:00 | 10125 | 0.67 |
| 25/08/2023 | 09:09:00 | 3225 | 0.67 |
| 25/08/2023 | 09:08:59 | 5225 | 0.67 |
| 28/08/2023 | 20:25:59 | 10125 | 0.65 |
| 29/08/2023 | 00:07:00 | 3225 | 0.64 |
| 29/08/2023 | 00:07:00 | 5225 | 0.64 |
| 02/09/2023 | 03:46:00 | 5225 | 0.61 |
| 02/09/2023 | 03:45:59 | 10125 | 0.61 |
| 03/09/2023 | 09:53:59 | 5225 | 0.61 |
| 03/09/2023 | 09:54:00 | 10125 | 0.61 |
| 03/09/2023 | 16:29:59 | 3225 | 0.61 |
| 03/09/2023 | 17:35:59 | 5225 | 0.61 |
| 04/09/2023 | 03:06:59 | 10125 | 0.60 |
| 04/09/2023 | 04:55:59 | 5225 | 0.60 |
| 04/09/2023 | 06:24:00 | 3225 | 0.60 |
| 04/09/2023 | 06:24:00 | 5225 | 0.60 |
| 04/09/2023 | 06:23:59 | 10125 | 0.60 |
| 06/09/2023 | 20:46:00 | 5225 | 0.58 |
| 06/09/2023 | 22:33:59 | 6875 | 0.58 |
| 06/09/2023 | 22:33:59 | 10125 | 0.58 |
| 07/09/2023 | 03:24:00 | 5225 | 0.58 |
| 07/09/2023 | 03:24:00 | 10125 | 0.58 |
| 07/09/2023 | 15:41:59 | 3225 | 0.58 |
| 07/09/2023 | 15:41:59 | 5225 | 0.58 |
| 07/09/2023 | 15:41:59 | 6875 | 0.58 |
| 07/09/2023 | 15:42:00 | 10125 | 0.58 |
| 07/09/2023 | 15:47:00 | 5225 | 0.58 |
| 07/09/2023 | 18:33:59 | 5225 | 0.58 |
| 07/09/2023 | 18:33:59 | 6875 | 0.58 |
| 08/09/2023 | 07:34:59 | 3225 | 0.57 |
| 08/09/2023 | 07:34:59 | 5225 | 0.57 |
| 08/09/2023 | 07:35:00 | 10125 | 0.57 |
| 14/09/2023 | 09:21:00 | 3225 | 0.52 |

**Table A3** *(continued)*

| date (UT) | time (UT) | Frequency (kHz) | Radial distance (au) |
|-----------|-----------|-----------------|----------------------|
| 14/09/2023 | 09:21:00 | 5225 | 0.52 |
| 14/09/2023 | 09:21:00 | 6875 | 0.52 |
| 14/09/2023 | 09:21:00 | 10125 | 0.52 |
| 15/09/2023 | 07:35:00 | 3225 | 0.51 |
| 15/09/2023 | 07:35:00 | 5225 | 0.51 |
| 16/09/2023 | 09:02:59 | 5225 | 0.50 |
| 16/09/2023 | 09:03:00 | 6875 | 0.50 |
| 16/09/2023 | 09:03:00 | 12225 | 0.50 |
| 18/09/2023 | 12:13:13 | 5225 | 0.48 |
| 18/09/2023 | 17:22:00 | 3225 | 0.48 |
| 18/09/2023 | 17:21:59 | 5225 | 0.48 |
| 18/09/2023 | 17:21:59 | 6875 | 0.48 |
| 18/09/2023 | 17:21:59 | 12225 | 0.48 |
| 18/09/2023 | 18:05:59 | 3225 | 0.48 |
| 18/09/2023 | 18:06:00 | 6875 | 0.48 |
| 19/09/2023 | 11:15:59 | 3225 | 0.46 |
| 23/09/2023 | 09:11:59 | 3225 | 0.41 |
| 23/09/2023 | 09:11:59 | 6875 | 0.41 |
| 23/09/2023 | 09:11:59 | 10125 | 0.41 |
| 24/09/2023 | 05:55:00 | 3225 | 0.40 |
| 24/09/2023 | 05:55:00 | 5225 | 0.40 |
| 24/09/2023 | 05:55:00 | 6875 | 0.40 |
| 24/09/2023 | 08:02:59 | 3225 | 0.40 |
| 24/09/2023 | 08:02:59 | 5225 | 0.40 |
| 24/09/2023 | 08:02:59 | 6875 | 0.40 |
| 24/09/2023 | 10:23:00 | 3225 | 0.40 |
| 24/09/2023 | 10:22:59 | 5225 | 0.40 |
| 24/09/2023 | 10:22:59 | 6875 | 0.40 |
| 24/09/2023 | 10:22:59 | 10125 | 0.40 |
| 24/09/2023 | 15:32:59 | 6875 | 0.40 |
| 25/09/2023 | 15:43:59 | 3225 | 0.38 |
| 28/09/2023 | 03:55:59 | 10125 | 0.32 |
| 28/09/2023 | 05:14:00 | 5225 | 0.32 |
| 28/09/2023 | 05:14:00 | 6875 | 0.32 |
| 03/10/2023 | 19:05:59 | 5225 | 0.20 |
| 03/10/2023 | 19:05:59 | 6875 | 0.20 |
| 03/10/2023 | 19:06:00 | 10125 | 0.20 |
| 03/10/2023 | 20:40:59 | 5225 | 0.20 |
| 04/10/2023 | 14:10:59 | 5225 | 0.17 |
| 04/10/2023 | 14:11:00 | 10125 | 0.17 |

**Table A3** *(continued)*

| date (UT) | time (UT) | Frequency (kHz) | Radial distance (au) |
|---|---|---|---|
| 04/10/2023 | 14:47:59 | 3225 | 0.17 |
| 04/10/2023 | 14:47:59 | 5225 | 0.17 |
| 04/10/2023 | 20:12:59 | 3225 | 0.17 |
| 04/10/2023 | 20:12:59 | 10125 | 0.17 |
| 06/10/2023 | 20:59:00 | 3225 | 0.11 |
| 06/10/2023 | 20:59:00 | 6875 | 0.11 |
| 06/10/2023 | 20:59:00 | 10125 | 0.11 |
| 07/10/2023 | 12:21:00 | 6875 | 0.08 |
| 09/10/2023 | 04:46:59 | 6875 | 0.01 |
| 10/10/2023 | 03:47:59 | 6875 | 0.01 |
| 10/10/2023 | 03:47:59 | 10125 | 0.01 |
| 13/10/2023 | 10:37:00 | 3225 | 0.11 |
| 13/10/2023 | 10:36:59 | 5225 | 0.11 |
| 13/10/2023 | 10:36:59 | 6875 | 0.11 |
| 19/10/2023 | 04:40:00 | 3225 | 0.28 |
| 19/10/2023 | 04:39:59 | 5225 | 0.28 |
| 19/10/2023 | 04:39:59 | 6875 | 0.28 |
| 19/10/2023 | 04:39:59 | 10125 | 0.28 |
| 29/10/2023 | 22:18:00 | 3225 | 0.48 |
| 29/10/2023 | 22:18:00 | 5225 | 0.48 |
| 29/10/2023 | 22:17:59 | 10125 | 0.48 |
| 30/10/2023 | 10:48:00 | 6875 | 0.50 |
| 30/10/2023 | 10:47:59 | 10125 | 0.50 |
| 30/10/2023 | 10:47:59 | 12225 | 0.50 |
| 30/10/2023 | 17:19:00 | 3225 | 0.50 |
| 30/10/2023 | 17:19:00 | 5225 | 0.50 |
| 30/10/2023 | 17:18:59 | 6875 | 0.50 |
| 30/10/2023 | 17:18:59 | 10125 | 0.50 |
| 30/10/2023 | 19:41:00 | 3225 | 0.50 |
| 30/10/2023 | 19:41:00 | 5225 | 0.50 |
| 30/10/2023 | 19:40:59 | 6875 | 0.50 |
| 30/10/2023 | 19:40:59 | 10125 | 0.50 |
| 31/10/2023 | 06:18:00 | 5225 | 0.51 |
| 31/10/2023 | 06:18:00 | 12225 | 0.51 |
| 31/10/2023 | 10:58:59 | 10125 | 0.51 |
| 31/10/2023 | 10:58:59 | 12225 | 0.51 |
| 01/11/2023 | 00:46:00 | 6875 | 0.53 |
| 01/11/2023 | 00:46:00 | 10125 | 0.53 |
| 02/11/2023 | 14:15:59 | 5225 | 0.54 |
| 03/11/2023 | 01:39:00 | 5225 | 0.56 |

**Table A3** *(continued)*

| date (UT) | time (UT) | Frequency (kHz) | Radial distance (au) |
|-----------|-----------|-----------------|----------------------|
| 04/11/2023 | 08:57:00 | 5225 | 0.57 |
| 04/11/2023 | 09:38:00 | 5225 | 0.57 |
| 04/11/2023 | 09:38:00 | 6875 | 0.57 |
| 04/11/2023 | 10:44:59 | 5225 | 0.57 |
| 04/11/2023 | 11:48:00 | 3225 | 0.57 |
| 04/11/2023 | 11:48:00 | 5225 | 0.57 |
| 04/11/2023 | 11:48:00 | 6875 | 0.57 |
| 14/11/2023 | 03:32:59 | 3225 | 0.69 |
| 14/11/2023 | 03:32:59 | 5225 | 0.69 |
| 14/11/2023 | 03:32:59 | 6875 | 0.69 |
| 14/11/2023 | 03:32:59 | 10125 | 0.69 |
| 23/11/2023 | 21:01:59 | 3225 | 0.77 |
| 23/11/2023 | 21:59:00 | 5225 | 0.77 |
| 23/11/2023 | 21:58:59 | 6875 | 0.77 |
| 23/11/2023 | 21:58:59 | 10125 | 0.77 |
| 23/11/2023 | 22:49:00 | 5225 | 0.77 |
| 23/11/2023 | 22:49:00 | 12225 | 0.77 |
| 24/11/2023 | 07:49:00 | 5225 | 0.78 |
| 24/11/2023 | 19:42:00 | 3225 | 0.78 |
| 24/11/2023 | 19:41:59 | 5225 | 0.78 |
| 24/11/2023 | 19:41:59 | 6875 | 0.78 |
| 24/11/2023 | 19:41:59 | 10125 | 0.78 |
| 24/11/2023 | 20:17:59 | 5225 | 0.78 |
| 25/11/2023 | 22:58:00 | 10125 | 0.79 |
| 25/11/2023 | 22:58:00 | 12225 | 0.79 |
| 26/11/2023 | 19:04:59 | 3225 | 0.79 |
| 26/11/2023 | 19:04:59 | 5225 | 0.79 |
| 26/11/2023 | 20:00:59 | 5225 | 0.79 |
| 26/11/2023 | 20:00:59 | 6875 | 0.79 |
| 26/11/2023 | 20:00:59 | 10125 | 0.79 |
| 26/11/2023 | 20:00:59 | 12225 | 0.79 |
| 28/11/2023 | 11:53:59 | 10125 | 0.81 |
| 30/11/2023 | 00:40:00 | 3225 | 0.82 |
| 30/11/2023 | 00:40:00 | 5225 | 0.82 |
| 30/11/2023 | 00:39:59 | 12225 | 0.82 |
| 01/12/2023 | 05:58:00 | 6875 | 0.83 |
| 01/12/2023 | 06:30:00 | 5225 | 0.83 |
| 01/12/2023 | 09:37:59 | 3225 | 0.83 |
| 01/12/2023 | 09:37:59 | 5225 | 0.83 |
| 01/12/2023 | 09:37:59 | 6875 | 0.83 |

**Table A3** *(continued)*

| date (UT) | time (UT) | Frequency (kHz) | Radial distance (au) |
|---|---|---|---|
| 02/12/2023 | 03:36:59 | 6875 | 0.83 |
| 02/12/2023 | 03:36:59 | 10125 | 0.83 |
| 02/12/2023 | 03:37:00 | 12225 | 0.83 |
| 09/12/2023 | 19:02:59 | 3225 | 0.87 |
| 09/12/2023 | 19:02:59 | 5225 | 0.87 |
| 12/12/2023 | 03:36:00 | 5225 | 0.88 |
| 12/12/2023 | 03:35:59 | 6875 | 0.88 |
| 12/12/2023 | 03:35:59 | 10125 | 0.88 |
| 12/12/2023 | 03:35:59 | 12225 | 0.88 |
| 18/12/2023 | 14:20:59 | 3225 | 0.89 |
| 18/12/2023 | 14:20:59 | 5225 | 0.89 |
| 18/12/2023 | 14:20:59 | 6875 | 0.89 |
| 18/12/2023 | 14:21:00 | 12225 | 0.89 |
| 21/12/2023 | 02:24:00 | 5225 | 0.90 |
| 21/12/2023 | 03:51:00 | 3225 | 0.90 |
| 21/12/2023 | 03:50:59 | 6875 | 0.90 |
| 21/12/2023 | 22:28:00 | 3225 | 0.90 |
| 26/12/2023 | 04:06:59 | 3225 | 0.90 |
| 26/12/2023 | 04:07:00 | 5225 | 0.90 |
| 26/12/2023 | 04:07:00 | 6875 | 0.90 |
| 26/12/2023 | 04:07:00 | 10125 | 0.90 |
| 26/12/2023 | 04:07:00 | 12225 | 0.90 |
| 26/12/2023 | 09:34:59 | 3225 | 0.90 |
| 26/12/2023 | 09:34:59 | 5225 | 0.90 |
| 02/01/2024 | 10:21:59 | 3225 | 0.90 |
| 02/01/2024 | 10:21:59 | 5225 | 0.90 |
| 02/01/2024 | 10:21:59 | 6875 | 0.90 |
| 02/01/2024 | 10:22:00 | 10125 | 0.90 |
| 02/01/2024 | 10:22:00 | 12225 | 0.90 |
| 04/01/2024 | 08:00:00 | 3225 | 0.90 |
| 04/01/2024 | 08:00:00 | 5225 | 0.90 |
| 04/01/2024 | 08:00:00 | 6875 | 0.90 |
| 04/01/2024 | 08:00:00 | 10125 | 0.90 |
| 04/01/2024 | 08:00:00 | 12225 | 0.90 |
| 04/01/2024 | 17:24:00 | 10125 | 0.90 |
| 04/01/2024 | 17:24:00 | 12225 | 0.90 |
| 05/01/2024 | 08:39:59 | 6875 | 0.90 |
| 05/01/2024 | 19:01:00 | 3225 | 0.90 |
| 05/01/2024 | 19:00:59 | 5225 | 0.90 |
| 05/01/2024 | 19:00:59 | 6875 | 0.90 |

**Table A3** *(continued)*

| date (UT) | time (UT) | Frequency (kHz) | Radial distance (au) |
|-----------|-----------|-----------------|----------------------|
| 05/01/2024 | 19:00:59 | 12225 | 0.90 |
| 06/01/2024 | 02:30:59 | 3225 | 0.89 |
| 06/01/2024 | 02:30:59 | 5225 | 0.89 |
| 06/01/2024 | 02:31:00 | 6875 | 0.89 |
| 06/01/2024 | 02:31:00 | 10125 | 0.89 |
| 06/01/2024 | 02:31:00 | 12225 | 0.89 |
| 12/01/2024 | 11:34:00 | 3225 | 0.88 |
| 12/01/2024 | 11:34:00 | 5225 | 0.88 |
| 12/01/2024 | 11:34:00 | 6875 | 0.88 |
| 12/01/2024 | 17:06:00 | 5225 | 0.88 |
| 13/01/2024 | 12:02:59 | 3225 | 0.88 |

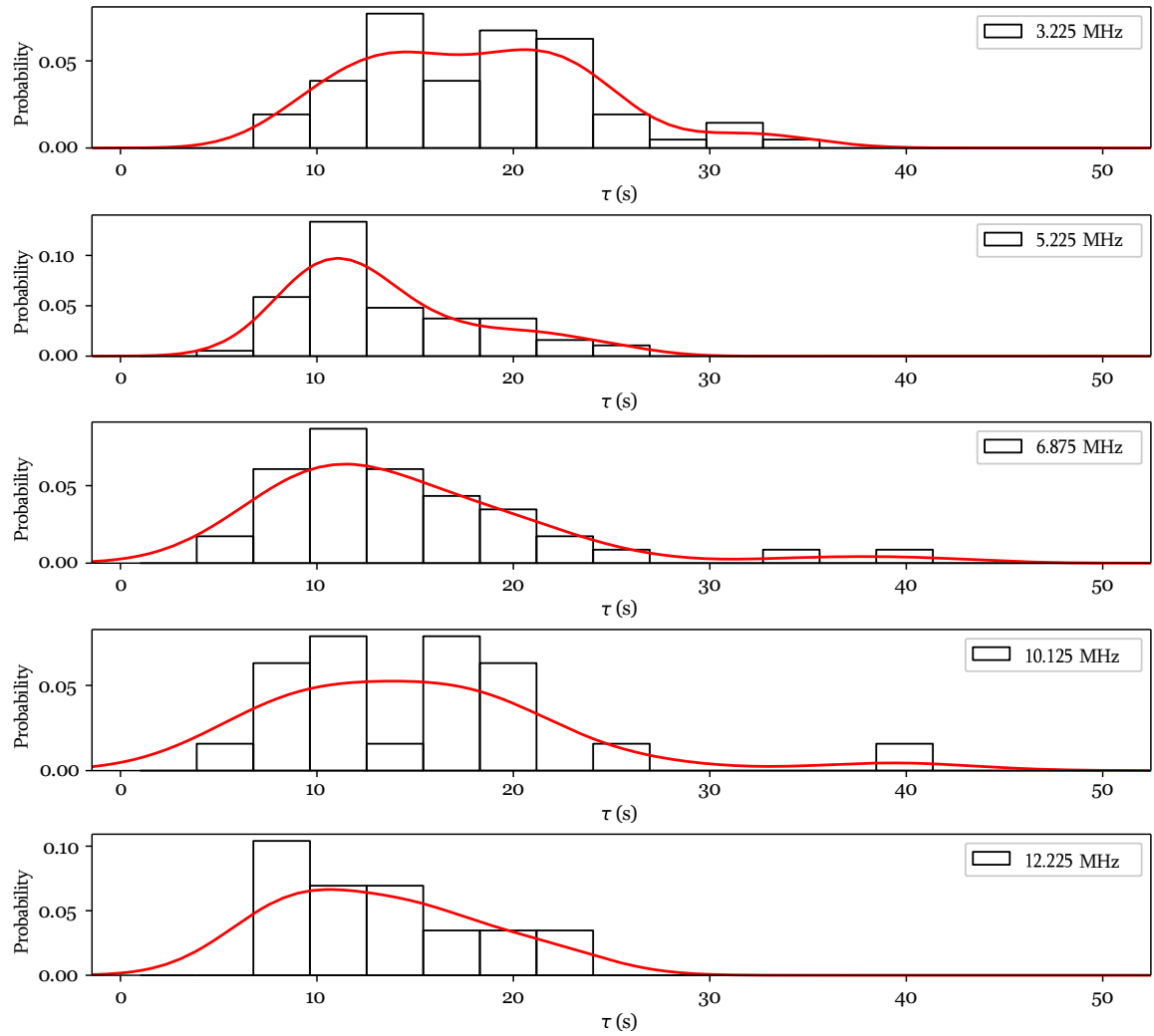

Figure A.5. Histograms of the decay times obtained from the visually selected type III burst subset at the five considered frequencies when the time resolution has been reduced to 3.5 s. Red full lines represent the empirical probability density function estimated using the KDE approach and constructed from the histogram.

## B. ADDITIONAL TEST TO CHECK THE EFFECT OF TIME RESOLUTION ON THE DECAY TIME MEASUREMENT

To further verify how the different time resolution of the data affect τ calculation we built up a set of synthetic light curves. These are derived as a power-law trend between the maximum and the background values as from Figure 4 of Krupar et al. (2020) at 5 MHz with a random noise with 0.5 dB standard deviation superimposed. The 0.5 dB uncertainty corresponds to the STEREO/Waves uncertainty (Krupar et al. 2012). Figure B.1 shows the results of the power-law fit on the synthetic light curves when different time resolutions are used. From the analysis, it appears that the power-law exponent and the corresponding error is a function of the is a function of the time resolution of the sample. The trend shows that with lower temporal resolution, shorter decay time are obtained.

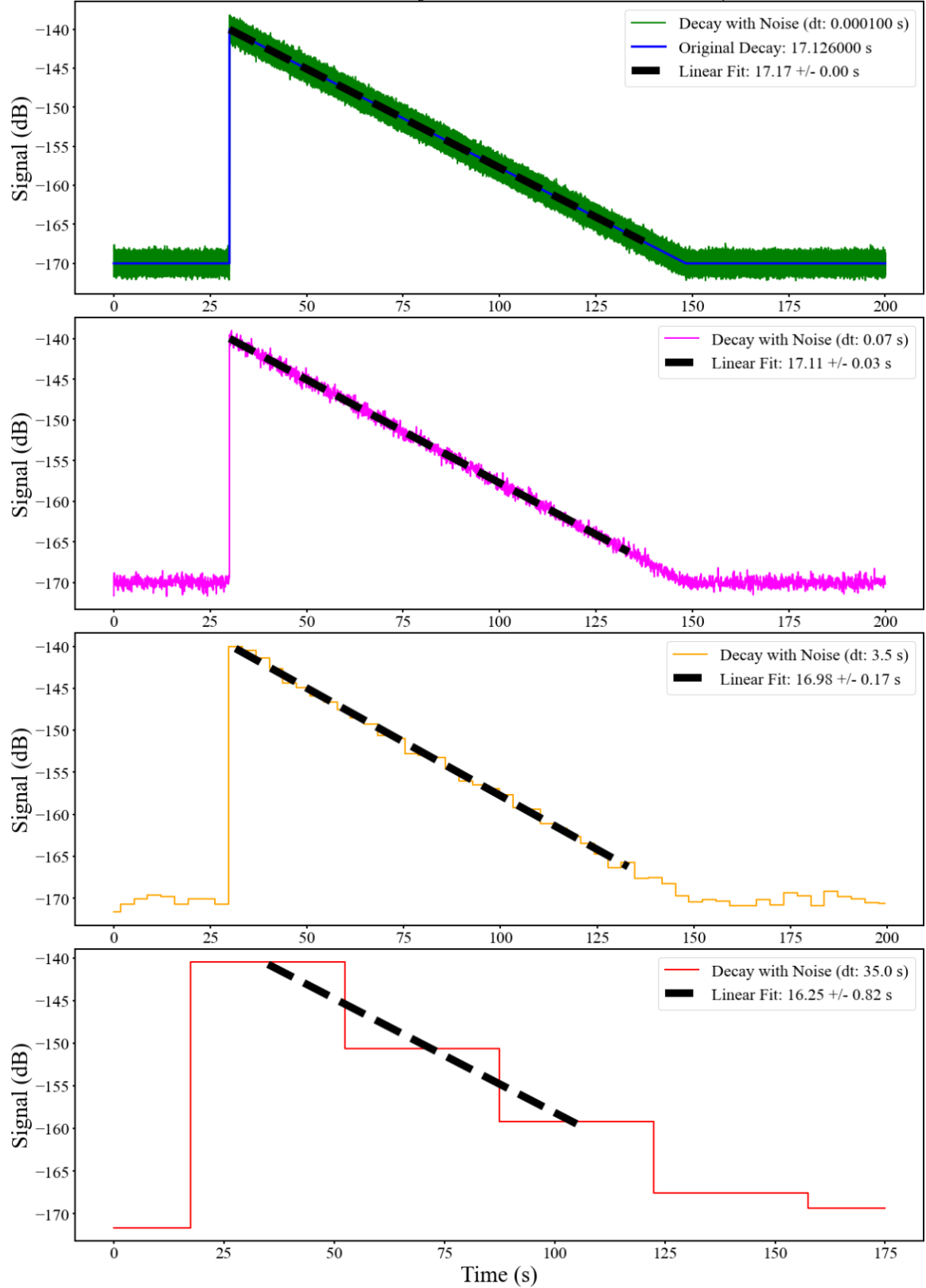

Figure B.1. Synthetic light curves at 5 MHz derived as a power-law functions between the maximum and the background values as from Figure 4 of Krupar et al. (2020) with a random noise with 0.5 dB standard deviation superimposed. Black dashed line corresponds to a power low fit on these data.